%% file: ms.tex
\documentclass[apjl]{emulateapj}
\usepackage{graphics}
\usepackage{apjfonts}
\usepackage{mathptmx}

\newcommand{\teff}{$T_{\!\mbox{\tiny\em eff}}$}

\newcommand{\teffq}{$T_{\!\mbox{\scriptsize \em eff}}^4$}

\newcommand{\hi}{H\,{\sc i}\rm}
\newcommand{\hii}{H\,{\sc ii}\rm}
\newcommand{\hei}{He\,{\sc i}\rm}
\newcommand{\heii}{He\,{\sc ii}\rm}

\newcommand{\caii}{Ca\,{\sc ii}}
\newcommand{\cai}{Ca\,{\sc i}}
\newcommand{\siii}{Si\,{\sc ii}}
\newcommand{\siiii}{Si\,{\sc iii}}
\newcommand{\siiv}{Si\,{\sc iv}}
\newcommand{\oi}{O\,{\sc i}}
\newcommand{\oii}{O\,{\sc ii}}
\newcommand{\nii}{N\,{\sc ii}}
\newcommand{\niii}{N\,{\sc iii}}
\newcommand{\cii}{C\,{\sc ii}}
\newcommand{\ciii}{C\,{\sc iii}}
\newcommand{\mgii}{Mg\,{\sc ii}}
\newcommand{\fei}{Fe\,{\sc i}}
\newcommand{\feii}{Fe\,{\sc ii}}
\newcommand{\tiii}{Ti\,{\sc ii}}
\newcommand{\crii}{Cr\,{\sc ii}}

\newcommand{\lin}{$\,\lambda$}
\newcommand{\llin}{$\,\lambda\lambda$}
\newcommand{\fglr}{\sc fglr\rm}
\slugcomment{}

\shorttitle{Blue supergiants in WLM}
\shortauthors{Bresolin et al.}

\begin{document}


\title{The Araucaria Project. VLT spectra of blue supergiants in WLM$^1$: classification and first abundances\\[2mm]}\footnotetext[1]{Based
on VLT observations for ESO Large Programme 171.D-0004.}

\author{Fabio Bresolin} \affil{Institute for Astronomy, 2680 Woodlawn
Drive, Honolulu, HI 96822; bresolin@ifa.hawaii.edu}

\author{Grzegorz Pietrzy\'nski} \affil{Universidad de Concepci\'on, Departamento de F\'isica, Casilla 160-C, Concepci\'on, Chile;  pietrzyn@hubble.cfm.udec.cl}

\author{Miguel A.~Urbaneja} \affil{Institute for Astronomy, 2680 Woodlawn
Drive, Honolulu, HI 96822; urbaneja@ifa.hawaii.edu}

\author{Wolfgang Gieren} \affil{Universidad de Concepci\'on, Departamento de F\'isica, Casilla 160-C, Concepci\'on, Chile; wgieren@coma.cfm.udec.cl}

\author{Rolf-Peter Kudritzki} \affil{Institute for Astronomy, 2680 Woodlawn
Drive, Honolulu, HI 96822; kud@ifa.hawaii.edu}

\and

\author{Kim A. Venn} \affil{Department of Physics \& Astronomy, University of Victoria, 3800 Finnerty Road, Victoria, BC V8P 1A1, Canada; venn@clare.phys.uvic.ca}

\begin{abstract}
As part of the Araucaria Project, we present the first spectral catalog of supergiant stars in the Local Group dwarf irregular galaxy WLM. In assigning a spectral classification to these stars we accounted for the 
low metal content of WLM relative to the galactic standards used in the MK process, by using 
classification criteria developed for B and A supergiants contained in the Small Magellanic Cloud. 
Our spectral catalog shows that our higher S/N spectroscopic sample of 19 objects contains at least 6 early-B (B0-B5) supergiants and 6 late-B and early-A (B8-A2) stars of luminosity class between Ia and II, as well as an O7~V star and an O9.7~Ia star. The spectra of several of these stars is of sufficiently high quality for a determination of the stellar parameters and abundances.
We have acquired also a second set of lower S/N spectra for mostly BA stars, however their quality 
does not allow a further analysis. We have carried out a quantitative analysis for three early-B supergiants. 
The mean oxygen abundance we derive is 12\,+\,log(O/H)\,=\,7.83 $\pm$ 0.12. 
This value agrees very well with the measurement that is obtained from \hii\/ regions. 
We therefore find no additional evidence for the discrepancy between stellar and nebular oxygen abundances measured for a single A-type supergiant by Venn et al. The analysis of B- and A-type supergiants yields compatible  results  for nitrogen, silicon and magnesium.
We show that the photometric variability of the blue supergiants included in our spectroscopic sample is negligible
for the use of these stars as distance indicators.

\end{abstract}

 

\section{Introduction}

Blue supergiant stars can be observed spectroscopically with current telescopes at distances of a few Mpc, reaching outside of the Local Group.
This has been possible only since recent times (\citealt{bresolin01}), allowing
to collect information on the physical properties and the chemical compositions 
of large samples of massive stars in different galactic environments. 
To achieve this goal, multi-object spectroscopy at moderate spectral resolution (\citealt{bresolin02}) provides an effective complement to the detailed spectral analysis of individual supergiants feasible from high-resolution spectra out to distances of $\sim$\,1~Mpc (\citealt{venn01}, \citealt{kaufer04}). 

A number of open issues linked to the investigation of the surface chemical composition of massive stars benefit from new  spectroscopic observations in the optical range of large samples of BA supergiants in nearby galaxies. These include the efforts to constrain mixing and wind properties in current models of stellar evolution  (\citealt{trundle05}), and to probe abundance gradients of various chemical elements in spiral galaxies in the context of galactic chemical evolution (\citealt{urbaneja05m33}).
In connection to the latter point, recently (\citealt{urbaneja05n300}) the analysis of blue supergiants has started to provide important checks on the oxygen abundances derived from \hii\/ regions in the central regions of spiral galaxies, where measurements of electron temperatures of the ionized gas suggest a major downward revision of the nebular metal abundances (\citealt{bresolin04,bresolin05}).

As part of the Araucaria Project (\citealt{gieren05}) we have obtained
spectra of blue supergiants in galaxies of the Local Group or its
immediate vicinities (WLM, NGC~6822, IC~1613, NGC~3109) and of the
Sculptor Group (NGC~55, NGC~247, NGC~300, NGC~7793) with the ESO VLT
telescope and its FORS spectrograph. The spectra obtained in NGC~300
(\citealt{bresolin02}) have led to the first determination of the
abundance gradient of $\alpha$-elements for a spiral galaxy located
beyond the Local Group (\citealt{urbaneja05n300}). Moreover, utilizing
the same data \citet{kudritzki03} showed that the stellar parameters
(\teff, log\,$g$) derived from the spectra of BA supergiants in
NGC~300 define a tight relation with the stellar luminosity: $M_{\rm
  bol} = a\log(g/$\teffq) + $b$.  This Flux-weighted
gravity--Luminosity Relationship (\fglr) is followed by supergiant
stars in a number of additional galaxies, and is being used in the
Araucaria Project, in combination with other stellar distance
indicators, to remove the potential effect that metallicity has on the
derivation of extragalactic distances from the Cepheid
Period-Luminosity relation.

With a total absolute magnitude $M_B\simeq- 14.0$, WLM is one of the faintest dwarf irregular galaxies in the Local Group. The distance to WLM has been measured by several authors using different stellar indicators. We will adopt the recent determination by \citet{mcconnachie05} based on the Tip of the Red Giant Branch (TRGB): $(m-M)_0=24.85\pm0.08$, corresponding to $932\pm33$ Kpc. The values measured by \citet{minniti97} via the TRGB method, 
by \citet{dolphin00} from a  fit to the color-magnitude diagram and by \citet{rejkuba00} from the 
magnitude of the horizontal branch are all consistent with the latest TRGB result.
The early Cepheid distance by \citet{sandage85}, who detected 15 of these variables,  suffered from an inaccurate
calibration of their photographic photometry (\citealt{ferraro89}). The Cepheid distance quoted by
\citet{lee93} is in agreement with the other techniques. Several new Cepheids have been discovered 
in the course of the Araucaria Project, and a distance based on the optical and near-infrared Period-Luminosity
relation is in preparation by our group. 
The stellar photometric studies mentioned above consistently derive a low value for the reddening across the
disk of WLM, with little indication for differential extinction. \citet{mcconnachie05} report $E(B-V)=0.035$, which is in fact the simple foreground component (\citealt{schlegel98}).

CCD photometry has shown that WLM possesses a young population concentrated in the disk of the
galaxy, superimposed on an old and extended metal-poor halo (\citealt{ferraro89}, \citealt{minniti96,minniti97}).
A metallicity of the old component [Fe/H]\,=\,$-1.45$ has been derived from the color of the Red Giant Branch (\citealt{minniti97}, \citealt{mcconnachie05}). Within the uncertainties, this is consistent with
the calculation of \citet{dolphin00} for the metallicity at the end of the initial episode of star formation
in WLM, derived from the color-magnitude diagram based on Hubble Space Telescope WFPC2 photometry.
The present-day metallicity has been measured from a few \hii\/ regions by \citet{skillman89}, \citet{hodge95} and
\citet{lee05}. These authors consistently obtained a low oxygen abundance, based on the electron temperature measured from the [O\,{\sc iii}]\lin4363 line in two of the brightest nebulae. The value reported by \citet{lee05}, 12\,+\,log(O/H)\,=\,$7.83\pm0.06$, corresponds to 15\% of the solar value [12\,+\,log(O/H)$_\odot$\,=\,8.66, \citealt{asplund04}). This low nebular oxygen abundance, [O/H]\,=\,$-0.83$ dex\footnote{Logarithmic abundance relative to the solar value: [O/H]$=$log(O/H) $-$ log(O/H)$_\odot$.}, 
while fitting the luminosity-metallicity relationship for dwarf irregular galaxies (\citealt{skillman89b}), contrasts with the result obtained for an A-type supergiant, [O/H]\,=\,$-0.21$ dex, by \citet{venn03}. This discrepancy between stellar and nebular chemical compositions is the first
of its kind so far detected in a dwarf irregular galaxy, and clearly warrants further investigation. An apparent mismatch between stellar and nebular abundances has been noted also in the spiral galaxy M31 (\citealt{trundle02}), but the \hii\/ region oxygen abundances in this galaxy are very uncertain, and the amount (or even existence) of the discrepancy 
heavily depends on which 
calibration for the strong-line abundance diagnostic $R_{23}$ (\citealt{pagel79}) is adopted. In WLM, on the other hand, the problem is more severe, because the nebular abundances are much more reliable, being based on the direct measurement of the electron temperature.

In this paper we focus on the brightest blue stars of WLM, centrally located in the disk of the galaxy, where recent low-level star formation has occured, as also indicated by the presence of gas photoionized by massive stars. We present spectroscopy of 38 stars distributed across the disk of WLM.
Several reasons make this galaxy an interesting target for a spectroscopic investigation of its blue supergiant content: (1) the spectral classification and the derivation of stellar parameters at low metallicity;
(2) the measurement of stellar chemical abundances, to be compared with the results from 
nebular studies; (3) the possibility of testing the \fglr\/ method at a significant distance of $\sim$1 Mpc.
In the current paper we address the first two points, by assigning a spectral classification to our
spectroscopic targets, and by measuring the chemical abundances of three early-B supergiants.
The investigation of the metallicity of late-B and early-A stars and of the \fglr\/ in WLM is deferred to a future paper, that will follow the publication of a stellar continuum-fitting technique we are developing.
In the current paper we discuss the observational data and their reduction in \S\,2. We present the spectral catalog for our sample in \S\,3. In \S\,4 we derive the chemical abundances for early-B supergiants, and compare them with the nebular abundances obtained in WLM. In \S\,5 we discuss the photometric variability of the spectroscopic targets, and conclude with a summary in \S\,6.

\begin{figure*}
\plotone{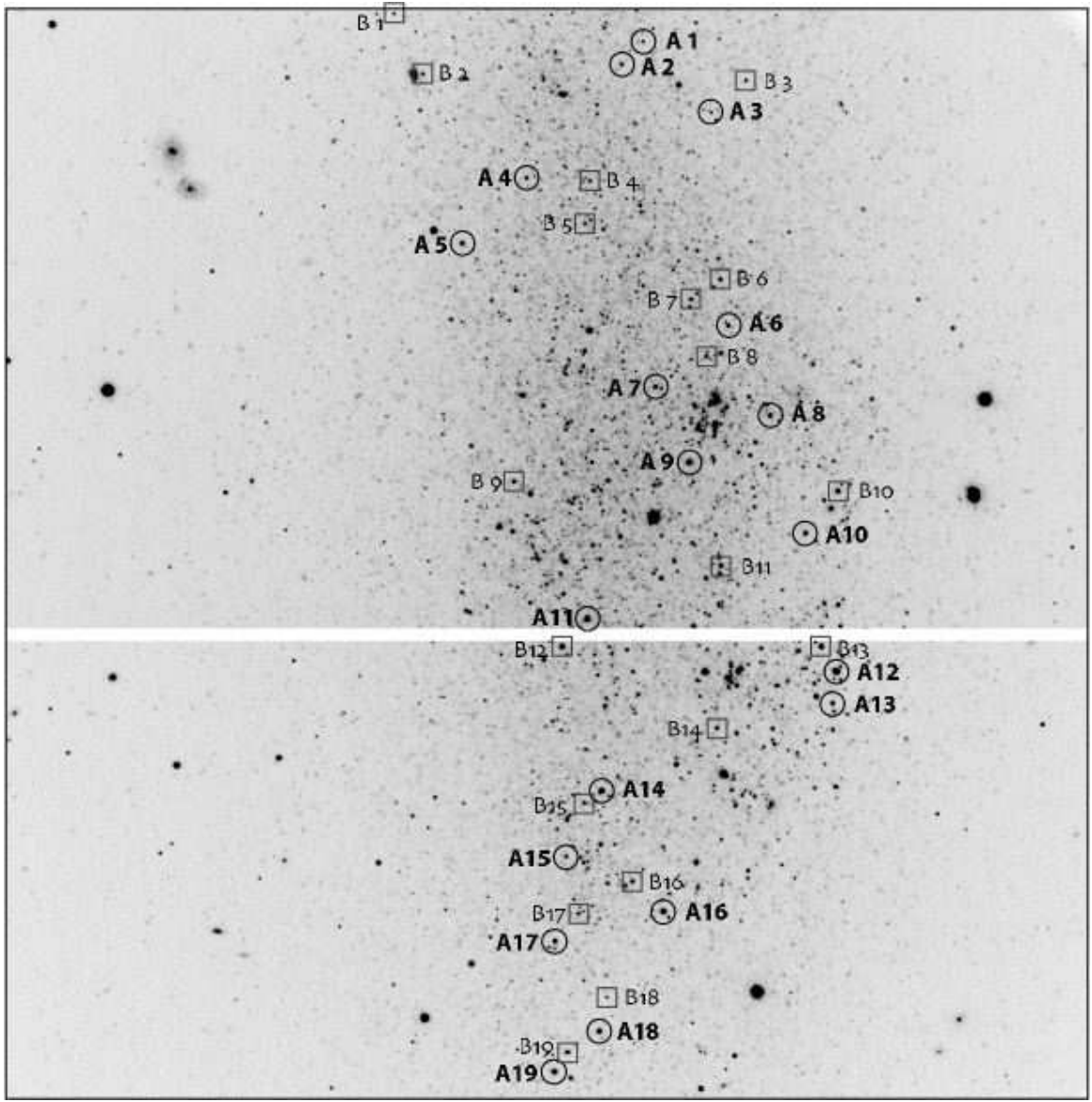}
\caption{We obtained multi-object spectroscopy of WLM supergiant candidates with two different slit setups. In this FORS2 image, used for these setups, we indicate 
with circles the targets observed in July 2003 (set A) and with squares
those observed in October 2003 (set B). The targets are numbered in progressive order, starting from the top of the image. The white horizontal band corresponds to
the gap in the FORS2 CCD mosaic, as projected on the sky. North is up and East is to the left.\label{map}}
\end{figure*}

\section{Observations}
Spectra of blue supergiant candidates in WLM have been acquired  at the Very Large Telescope UT4 (Yepun) with the Focal Reducer and low dispersion Spectrograph~2 (FORS2) in MOS mode. Candidates for this study were selected on the basis of their $V-I$ color index and $V$ magnitude, measured from the photometry obtained as part of our Cepheid variable search program at the Las Campanas Polish 1.3m telescope (Pietrzy\'nski et al.~2006, in preparation; see \citealt{pietrzynski02} for details on the data reduction). A large fraction of our targets are included in the \cite{sandage85} list of
bright blue stars (the SC numbers in Table~1). We also included in the sample the A-type supergiant SC\,15 analyzed by \citet{venn03}, as well as the early-B 
star SC\,35, for which these authors could not measure \teff, because of the unavailability of the 
\siiii\/ diagnostic in their spectrum.
We have compared the $V$ magnitudes of our targets with the photometry of \citet{mcconnachie05}. 
The agreement between 35 objects in common, in the order of 0.01 mag or better, with no significant zero-point offset, is excellent and in line with other comparisons 
published in our series of papers (e.g.~\citealt{pietrzynski02,pietrzynski04}).

\begin{figure}
\plotone{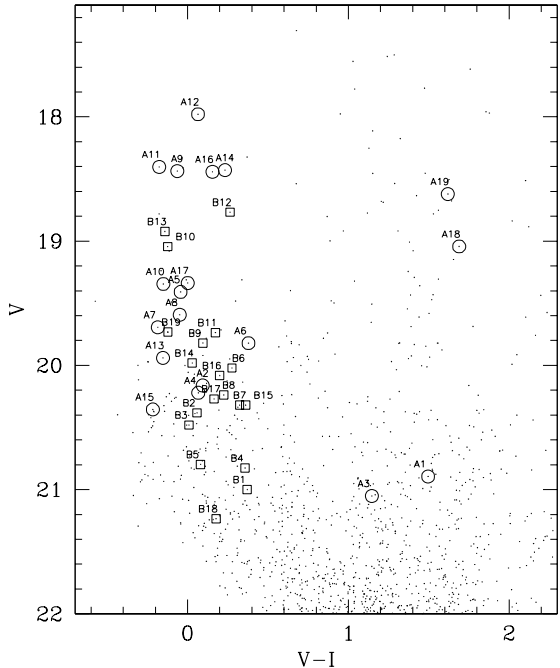}
\caption{The stars for which we have obtained spectra indicated on the $V~vs~V-I$ color-magnitude
diagram. 
}
\label{cm}
\end{figure}

The spectra were acquired on two separate nights: July 28 and October 26, 2003.
On both nights, we obtained spectra of 19 stars with the 600B grism, achieving 
a resolution of $\sim$\,5\,\AA. The airmass during the observations was
smaller than 1.07.
The spectra span approximately 2500~angstroms on the $2048\times4096$ FORS2 mosaic, centered around 4500~\AA.
The stellar targets for the two sets of observations (denoted with the letters A and B in the remainder of the paper for the July and October runs, respectively) 
are positioned in the same section of the galaxy, where young stars and \hii\/ regions are concentrated (see Fig.~\ref{map}). Our set B includes stars
that received lower priority during our target selection, either because of their
fainter magnitudes or because of the higher probability for the presence of neighboring stars, that might contaminate the spectra. We must also mention the fact that at the northern and southern extremes of the observed field
no suitable blue supergiant candidate was available. In these cases we obtained spectra of bright red stellar objects. The location of the spectroscopic targets in the $V~vs~V-I$ color-magnitude diagram is shown in Fig.~\ref{cm}.

The total exposure times were 4500 seconds (set A) and 3600 seconds (set B).
The seeing was good throughout the first night (0\farcs7), but only mediocre (1\farcs2) during the second. Correspondingly, the signal-to-noise ratio (S/N) for our second set of spectra is not sufficiently high (S/N\,=\,20--40 in most cases) to allow a quantitative analysis for an accurate derivation of the stellar parameters. 
For set A, on the other hand, the S/N per pixel at 4500~\AA\/ is larger than 40 for all but four stars, reaching a maximum of $\sim$120.

Although rectified spectra are generally sufficient for the derivation of chemical abundances and stellar parameters, we also decided to flux calibrate the stellar spectra. For this purpose the spectrophotometric 
standards Feige~110, CD-32~9927 and LTT~4816 were oberved. The spectral energy distributions thus obtained will be used, in conjunction with the line features and model spectra, to 
constrain the stellar parameters in a future paper on the \fglr.

The two frames available in each field were combined, in order to increase the S/N ratio. 
We extracted the individual 2-D spectra (extending $\sim$\,20 arcsec along the spatial direction), corresponding to each of the 19 slitlets, from the average
images and proceeded by utilizing common {\sc iraf} tasks for longslit spectroscopy data reduction. These included bias and flat-fielding corrections, cosmic ray rejection and wavelength calibration via He+HgCd lamp exposures. We finally obtained sky-subtracted 1-D spectra, normalized using a low-order polynomial. A flux-calibrated version of the extracted spectra was also created.\\

\section{Spectral catalog}
Because of the low metallicity of WLM classical MK spectral type indicators
for early-type stars need to be adjusted in order to maintain a meaningful
relation between spectral type and stellar physical properties as the strength of the metal lines becomes very low.
This is of course important when comparing the massive stellar content of different galaxies, and is crucial for the work carried out in the Araucaria Project, that comprises galaxies of vastly different metallicities, spanning roughly one order of magnitude in Fe/H.

A number of authors have considered the classification of stars that are metal deficient relative to the Galactic MK standards. While the classification of O stars relies essentially on the \heii-\hei\/ ionization equilibrium, and is therefore independent of metallicity, for B stars the classification relies on trends in the strength of metal lines, mostly from silicon and magnesium, and in the case of the later B-types these are measured in relation to \hei\/ features. \citet{fitzpatrick91} showed how to transfer the MK classification framework to the B supergiants in the moderately metal-poor environment of the Large Magellanic Cloud (LMC), where the strengths of the spectral metal features are reduced by $\sim$\,30\% relative to the Galactic standards.
Similarly, \citet{lennon97} considered the case of B supergiants in the Small
Magellanic Cloud (SMC), where the effects of the metal deficiency on the 
spectral classification are more important than in the case of the LMC.
A classification scheme for A supergiants in the SMC was more recently developed by
\citet{evans03}.

The weakness of the diagnostic lines in the WLM stellar spectra implies that good S/N is necessary for a reliable classification. In our case, this was achieved for most of the targets in set A (the typical S/N per pixel is larger than 50), while for set B in general we were only able
to provide a rough estimate of the spectral type. 
We also found that for some objects in the latter set of spectra the contamination due to the presence of unresolved companions appears to be significant. 

A further consideration about our spectroscopic data concerns their rather coarse
spectral resolution, about 5\,\AA, relative to the 1-2~\AA\/ resolution
attained in the Magellanic Cloud works mentioned above, as well as in
widely used digital spectral catalogs of early-type stars (e.g.~\citealt{walborn90}). The consequent blending affects some of the spectral features employed in the classification (e.g.~\siiv\,\lin4116 and \hei\,\lin4121 in early B stars). Moreover, the measurement of
weak metal line equivalent widths is not feasible at this resolution, preventing us from carrying out quantitative tests on the trends of metal line strenghts with 
spectral type.
However, overall we are confident about the reliability of the derived spectral classes, since in general these trends can be estimated qualitatively with good accuracy, provided the S/N of the spectra is sufficiently high.

\input{tab1}

\subsection{Classification criteria: spectral classes}

For the classification of B supergiants we adopted the criteria introduced by \cite{lennon97}, developed for supergiant spectra in the SMC, where the present-day metal content is comparable to that found for WLM. For A-type supergiants, we 
used instead the scheme introduced by \citet{evans03} in their study of SMC stars, i.e.~we relied on the strength of the \caii\/ H and K lines, namely the ratio
of the line depths \caii\/ K/(\caii\/ H + H$\epsilon$).
This classification method has the advantage of being based on strong metal lines, and can therefore be applied 
to spectra of low spectral resolution and low S/N.
The interstellar component of the \caii\/ lines, that cannot be disentangled from the stellar component at our spectral resolution, is expected to play only a
second-order effect on the measured ratio. 
We note the different philosophy behind these two classification criteria: while the B-supergiant
classification scheme was introduced by \citet{lennon97} in order to extend the relation between spectral type and
effective temperature observed in the Galactic standards to lower metallicities, the solution proposed for A-supergiants by \citet{evans03} allows for a metallicity dependence of the spectral type-temperature relation.

Our sample also includes two O stars. For their classification
we relied on the stellar atlas by \citet{walborn90}, supplemented by information
contained in \citet{walborn71} and \citet{walborn00}.

\subsection{Luminosity  classes}

The luminosity classification was based on the empirical relationship between 
the equivalent width of H$\gamma$ ($W_\gamma$) and the luminosity class derived  for O9-F8 stars in the SMC by \citet{azzopardi87}. However, the value of $W_\gamma$
for three A supergiants in set A exceeds the maximum value assigned to the lowest luminosity class (Ib) in Azzopardi's work. In these cases (stars A2, A4 and A14) we have therefore
assigned a luminosity class II. This matches
what one obtains using the luminosity criteria adopted by \citet{evans04} for A-type supergiants contained in their 2dF survey of the SMC (their Table~5). Most of the stars in set B, including two B-type stars, have $W_\gamma$ larger than for Ib supergiants, and were similarly classified as luminosity class II objects.

\begin{figure*}
\epsscale{0.9}
\plotone{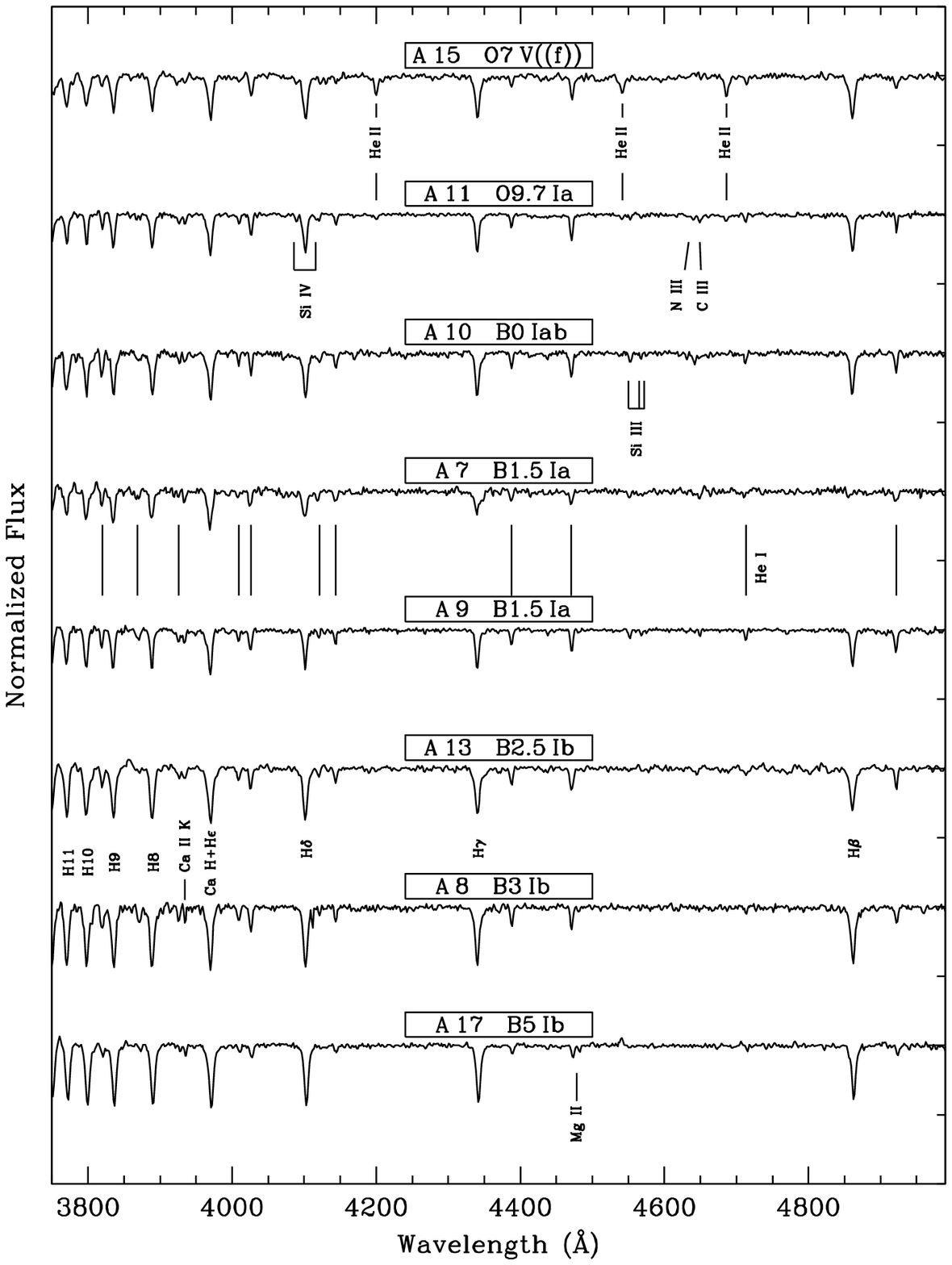}
\caption{Normalized spectra of O and early-B supergiants in set A.
The spectral features identified are:
{\em (below A15)} 
\heii\,\lin4200,\lin4542,\lin4686;
{\em (below A11)} 
\siiv\,\lin4089,\lin4116, \niii\,\llin4634,40-42, \ciii\,\lin4650;
{\em (below A10)} 
\siiii\,\llin4553-68-75;
{\em (below A7)} 
\hei\,\lin3820,\llin3867-72,\lin3926,\lin4009,\lin4026,\lin4121,\lin4144,\lin4388,\lin4471,\lin4713,\lin4922;
{\em (below A13)}
Balmer lines from H$\beta$ to H11, and the \caii\,K (\lin3933) and H (\lin3968) lines;
{\em (below A17)} 
\mgii\,\lin4481
}\label{spectraA1}
\end{figure*}
\epsscale{1.0}

\begin{figure*}
\epsscale{0.9}
\plotone{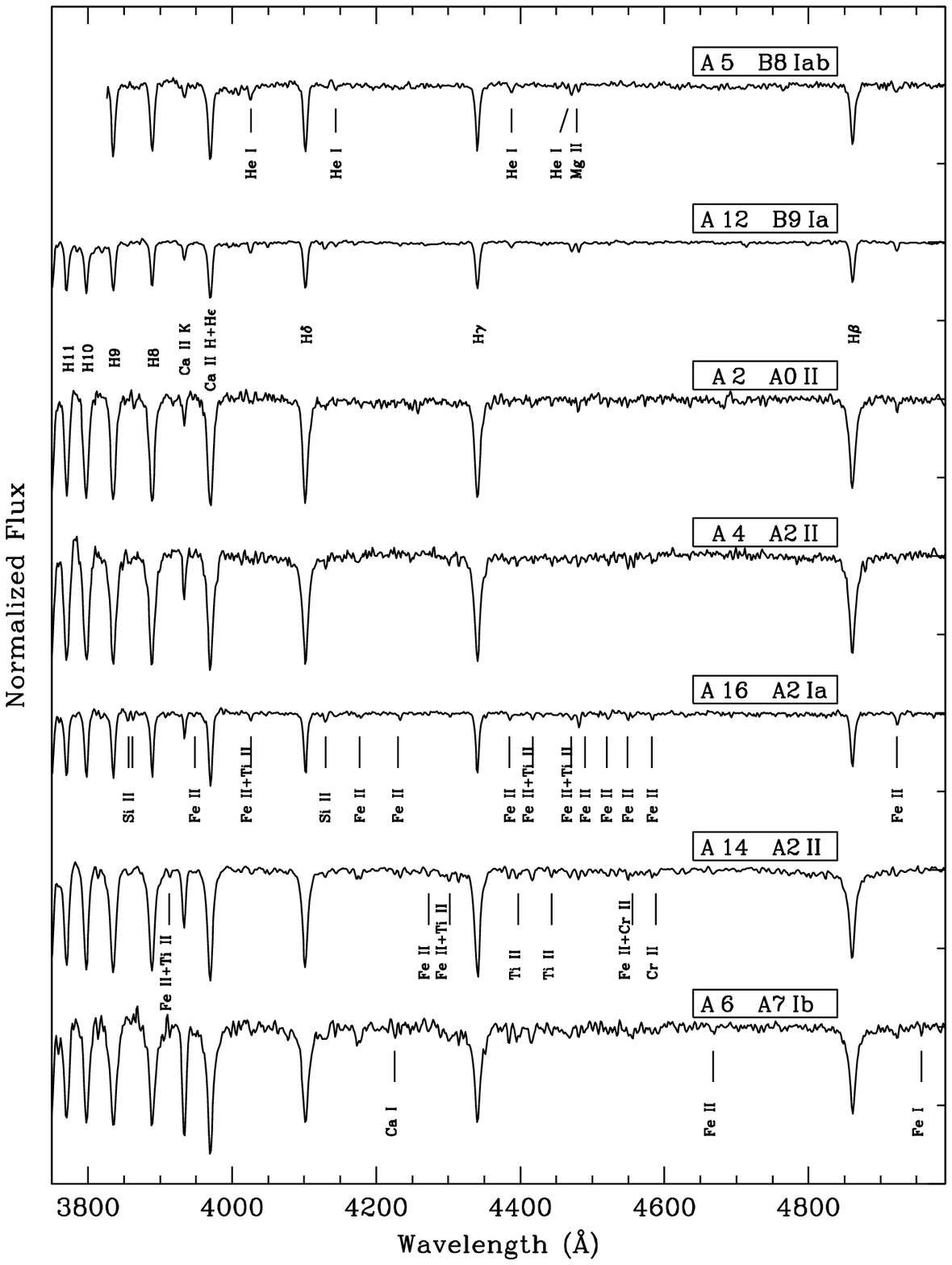}
\caption{\footnotesize Normalized spectra of late-B and A supergiants in set A. The
spectral features identified are: 
{\em (below A5)} 
\hei\,\lin4026,\lin4121,\lin4388,\lin4471, \mgii\,\lin4481; 
{\em (below A12)}
Balmer lines from H$\beta$ to H11, and the \caii\,K (\lin3933) and H (\lin3968) lines;
{\em (below A16)}
\siii\,\llin3856-62,\llin4128-32, 
\feii\,\lin3945,\lin4024,\llin4173-78,\lin4233,\lin4385,\lin4417,\lin4473,\llin4489-91,\lin4549,\lin4583,\lin4923,
\tiii\,\lin4028,\lin4418,\llin4469-71;
{\em (below A14)} \feii\,\lin3914,\lin4273,\lin4303,\lin4556,
\tiii\,\lin3913,\lin4300,\llin4395-99,\lin4444,
\crii\,\lin4588,\lin4559;
{\em (below A6)} \feii\,\llin4666-69, \fei\,\lin4957, \cai\,\lin4226
}\label{spectraA2}
\end{figure*}
\epsscale{1.0}

\begin{figure*}
\epsscale{0.9}
\plotone{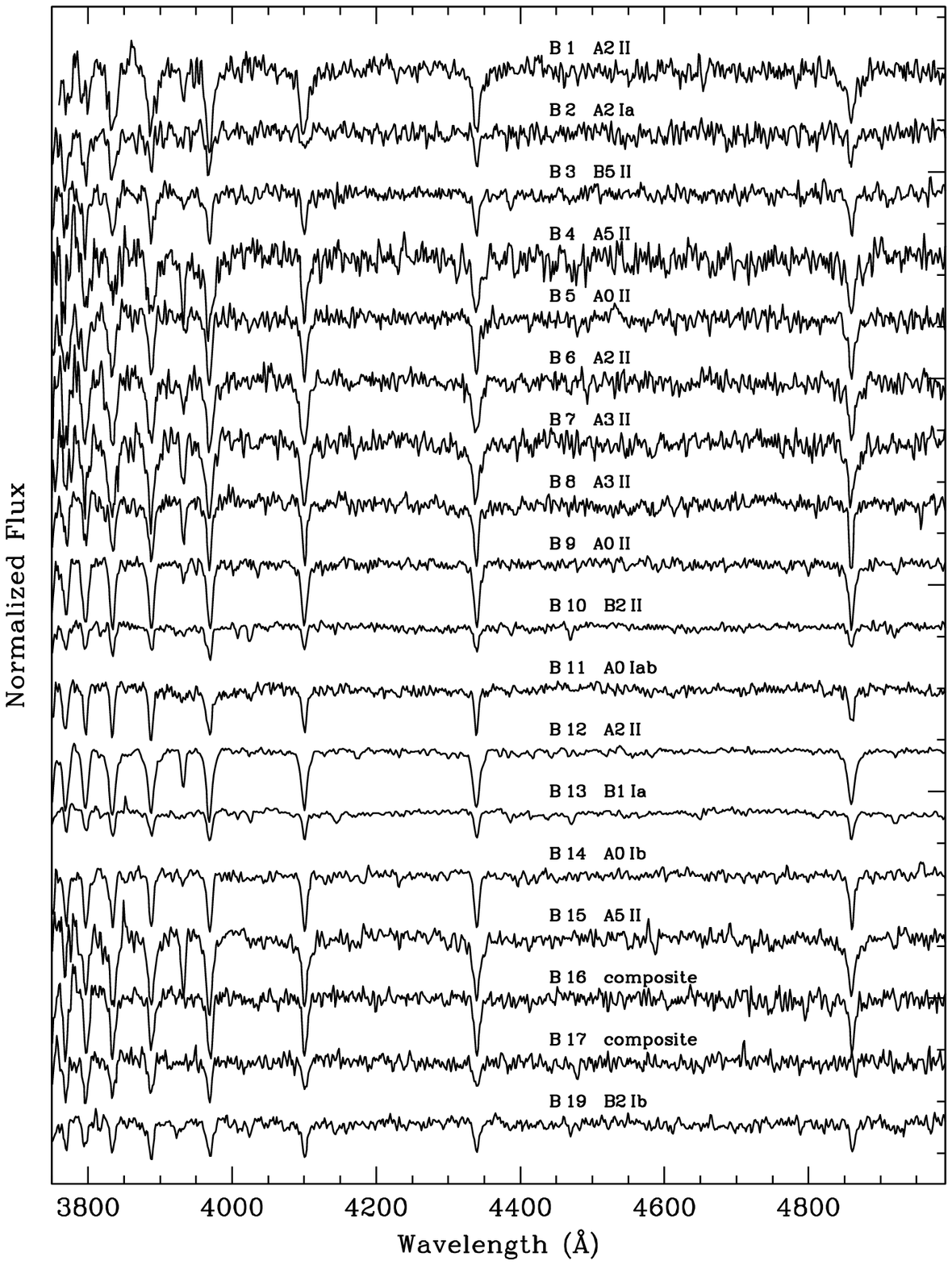}
\caption{Normalized spectra for set B. The dominant spectral type is indicated.
}\label{spectraB}
\end{figure*}
\epsscale{1.0}

\subsection{Results}

We were able to apply the criteria above for all stars in set A (excluding four late-type stars), but the quality of set B spectra is in many instances not sufficient for an accurate classification. 
In this case, we have either attempted to match the observed spectra with
SMC supergiant templates (B stars), admittedly producing a rather uncertain classification, or used the \caii\/ K/(\caii\/ H + H$\epsilon$)
criterion (A stars). We could not converge on a 
single dominant spectral type for three objects in set B (B16-B18), therefore they are labeled as 'composite' in Table~1, where we summarize 
our results. For each blue supergiant we include:
celestial coordinates (columns~2 and 3; measured with reference to the system defined by the USNO-2.0 catalog), photometric data ($V$ in col.~4 and $V-I$ in col.~5), the spectral type (col.~6), the ratio 
\caii\/ K/(\caii\/ H + H$\epsilon$) (col.~7), the equivalent width of H$\gamma$
($W_\gamma$, col.~8), the mean S/N per pixel of the spectrum (col.~9), and
the heliocentric velocity measured from the Balmer lines ($V_{\rm Helio}$, col.~10). The SC number from \citet{sandage85}, when available, is included in column 11.
Given the low resolution of the spectra, the uncertainty in $V_{\rm Helio}$ is around 10-15 km\,s$^{-1}$.
However, we note that the radial velocity information that we obtain across the galaxy is consistent with the
\hi\/ rotation curve measured by \citet{jackson04}, with objects in the northern half 
approaching us relative to the systemic velocity of $-122$ km\,s$^{-1}$ (\citealt{koribalski04}), and the southern half receding. 

In Fig.~\ref{spectraA1} and Fig.~\ref{spectraA2} we show the normalized spectra of all the blue supergiants in set A, organized by spectral type, and starting with the earliest types. The lower-quality spectra of set B are shown in Fig.~\ref{spectraB}. 

As we remarked above, the classification criteria adopted here for A-type supergiants allow for a variation
of stellar \teff\/ as a function of metallicity for a given spectral type, 
i.e.~the spectrum of a star of low metallicity could resemble that of a {\em hotter} star of higher 
metallicity, because of the reduced strength of the metal features. This
explains the different classification obtained for SC~15 and SC~31 by us (A2) and
\citet{venn03} (A5), since the latter authors prefer to assign spectral types based
on the measured temperature and a \teff-spectral type relation valid for Galactic A  supergiants.

We also note that the estimation of the stellar luminosity from the equivalent width of the Balmer lines can lead to apparently inconsistent results. This can be seen, for example, by comparing stars A14 and A16. These are both A2 stars, with virtually equal $V$ magnitudes ($V=18.4$), but the former is classified as a luminosity class II object, based on its large $W_\gamma=6.2$, while the latter is a class Ia star having $W_\gamma=2.4$. Another luminosity class II A2 star, our object A4 in Table~1, is more than 1.5 magnitudes fainter.
With the adopted value of the distance to WLM, and accounting for the small extinction towards it, stars A14 and A16 have an absolute magnitude $M_V\simeq-6.6$. According to the Galactic $M_V$-luminosity class calibration by \citet{humphreys84} these two stars would be classified as Iab supergiants, while A4 would be classified as Ib (other calibrations we considered would yield  the same results). We note that the standard deviation in the \citet{humphreys84} calibration is typically 0.5 magnitudes for a given luminosity class.
A large scatter in $W_\gamma$ within a given luminosity class at fixed spectral type has also been noted more recently by \citet{evans04}, who decided to 
use the information on the stellar brightness to define the luminosity class.
In our case, however, rather than trying to adjust our luminosity classes from the knowledge of the absolute magnitude of the stars we will keep basing our stellar classification on the spectral morphology alone.

Foreground contamination is expected to be negligible for all spectral types of interest here at the galactic latitude of WLM ($b=-73\fdg6$), as confirmed by the predictions from the Besan\c{c}on models (\citealt{robin03}). We have estimated that the contamination from foreground dwarfs and giants is 1.5\% in the region of the color-magnitude diagram occupied by the G stars A18 and A19, and 0.2\% for the fainter G stars A1 and A3. For blue stars (B and A types) enclosed in the region of the color-magnitude diagram $18<V<21$, $-0.4<(V-I)<0.4$ the calculated contamination is 0.01\%.\\

\subsection{Comments on individual stars}

We motivate here the detailed classification of the OB stars contained in our set A, based on the criteria used by \citet{lennon97} for SMC supergiants.
For A supergiants this is not necessary, as we derived the spectral classes simply from the \caii\/ K/(\caii\/ H + H$\epsilon$) ratio (Table~1):\\

{\em A5.--} The relative intensity of \hei\,\lin4471 and \mgii\,\lin4481
is the main temperature discriminant for late-B supergiants. The two lines 
have approximately the same strength at B8, and \mgii\,\lin4481 becomes
progressively stronger relative to \hei\,\lin4471 for B9 and early A types.
For this star, we observe \mgii\,\lin4481 $<$ \hei\,\lin4471. In addition,
\hei\,\lin4144 $>$ \hei\,\lin4121. These conditions apply to B8 stars
in the scheme by \citet{lennon97}.\\

{\em A7.--} We assigned a B1.5 class from the absence of \heii\,\lin4686 (present 
in the B0 and B1 types), the absence of \siiv\,\lin4116 and the presence of
\siiv\,\lin4089 $\simeq$ \oii\,\llin4072-76. We excluded types B2 or later 
since no \siiv\/ lines could be detected.
This star, together with the nearby A8, A9 and A10, lies in a diffuse nebulosity
surrounding the group of bright stars associated with the \hii\/ regions 
HM8 and HM9 (complex C1 of \citealt{hodge95}). The Balmer lines of these four stars are mildly
affected by weak nebular emission. This could explain the filling of the H$\beta$ line and the Ia luminosity class deduced from $W_\gamma$, despite the rather faint $V$ magnitude.\\

{\em A8.--} In the scheme by \citet{lennon97}, the \siiii\,\lin4553 $\sim$ 
\mgii\,\lin4481 at B2.5. At the cooler B3 type \siiii\,\lin4553 $<$ 
\mgii\,\lin4481, as observed in this star. The \siiii\/ line disappears at B5.\\

{\em A9.--} The B1.5 classification derives from the simultaneous absence of 
\siiv\,\lin4116 and presence of \siiv\,\lin4089 $\leq$ \oii\,\llin4072-76.
A weak nebular emission affects the Balmer lines. Detailed stellar parameters and the metal content for this star, as well as for A10 and A11, are measured in Sect.~\ref{sec_abundances}.\\

{\em A10.--} The \heii\/ lines at 4200, 4542 and 4686 \AA\/ are all present in 
this star's spectrum, while \heii\,\lin4542 $<$ \siiii\,\lin4553, corresponding to
a B0 type. The Balmer lines are affected by weak nebular emission.\\

{\em A11.--} The O9.7 type derives from the similarity in the strengths of
\heii\,\lin4542 and \siiii\,\lin4553 (\citealt{walborn71}). The ratios 
\heii\,\lin4200/\hei\,\lin4144 and \heii\,\lin4542/\hei\,\lin4388 are also
consistent with the adopted spectral class.\\

{\em A12.--} \citet{sandage85} identified this as the visually brightest star in WLM. Our CCD
magnitude, $V\,=\,17.98$, corresponds to $M_V\,\simeq\,-7.0$ when accounting for the foreground extinction. 
Note that this is more than 2 magnitudes fainter than the visually brightest blue stars (late-B to early-A supergiants, for which the bolometric corrections are small) found in more massive irregular galaxies, such as the Magellanic Clouds.
We assign a B9 class from the conditions \mgii\,\lin4481 $>$ \hei\,\lin4471 and \feii\,\lin4233 $<$ \siii\,\llin4128-32.\\

{\em A13.--} The \siiv\/ lines are absent in this star, implying a type later than
B1.5. Although the metal line strenghts are uncertain at the FORS resolution and with the S/N of the observed spectrum, the \siiii\,\lin4553 is comparable to \mgii\,\lin4481, which indicates a type close to B2.5. The presence of \siiii\,\lin4553 excludes a type later than B3.\\

{\em A15.--} The main spectral type indicator for O stars is the ratio of 
\hei\,\lin4471 to \heii\,\lin4542. For this star, the \hei\/ line is slightly stronger than
the \heii\/ line. This is also true for \hei\,\lin4026 relative to \heii\,\lin4200. 
The \siiv\,\lin4089 is well detected. These criteria  point to a O7 type, or perhaps
a slightly later type. If we use $\log W' = \log W(4471) - \log W(4542) \simeq 0.0$, we arrive at the same conclusion (\citealt{conti71}).
The weak \nii\,\llin4634,4640-42 emission is associated 
with strong \heii\/ absorption, therefore the ((f)) designation. The V luminosity
class derives from the strong \heii\,\lin4686 in absorption.\\

{\em A17.--} The \siiii\/ lines are absent (type $\geq$ B5), with \mgii\,\lin4481 $<$ \hei\,\lin4471 (type $\leq$ B8). The relative strength of \siii\,\llin4128-32 
and \hei\,\lin4121 is used to distinguish the B5 and B8 classes. However,
in our spectrum these two lines are very weak and have apparently the same
strength. Somewhat arbitrarily we assign a B5 classification from the
fact that the \hei\,\lin4471/\mgii\,\lin4481 ratio is larger than in
our other B8 star (A5).\\

{\em G-type stars.--} Four objects in Table~1 (not shown in Fig.~\ref{spectraA1}-\ref{spectraA2}) have been classified as G stars.
We have used the criteria in \citet{evans03}: G-band/$H_\gamma\simeq0.9$ for G0 (star A3), $H_\gamma>$\,\fei\,\lin4325 for G2 (A18, A19) and $H_\gamma\simeq$
\fei\,\lin4325 for G5 (A1). The negligible foreground contamination and the measured radial velocities, that
indicate that these stars participate to the galactic rotation (see Table~1),  
strongly support the association of these four stars  with WLM.\\

\epsscale{0.86}
\begin{figure*}[hb]
\plotone{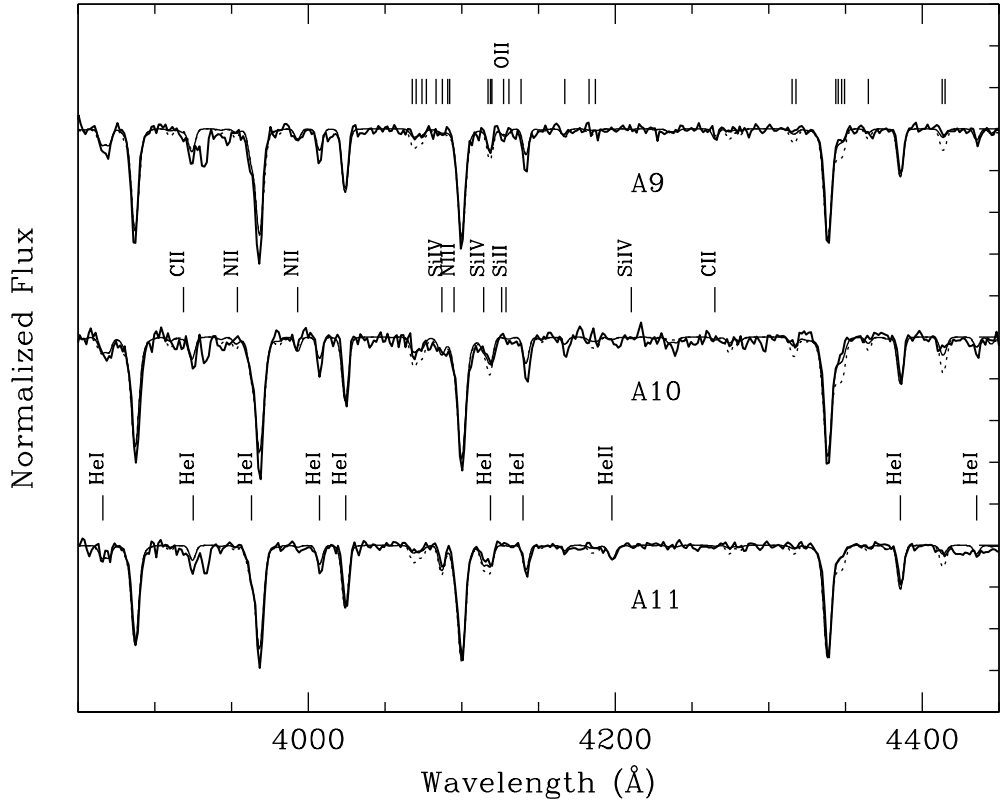}
\caption{Comparison between the observed spectrum of the B-type supergiants
A9, A10 and A11 (thick lines) with the best-fitting models (thin lines) in the
3850--4550~\AA\/ range. The dotted lines represent {\sc fastwind} models
calculated for a metallicity scaled up to the oxygen abundance measured
by \citet{venn03}, 12\,+\,log(O/H)\,=\,8.45. The line
identification provided for \oii\/ lines (top), additional metal lines (center) 
and He lines (bottom) refers to some of the strongest present.}\label{models1}
\end{figure*}
\epsscale{1.0}

\epsscale{0.86}
\begin{figure*}[ht]
\plotone{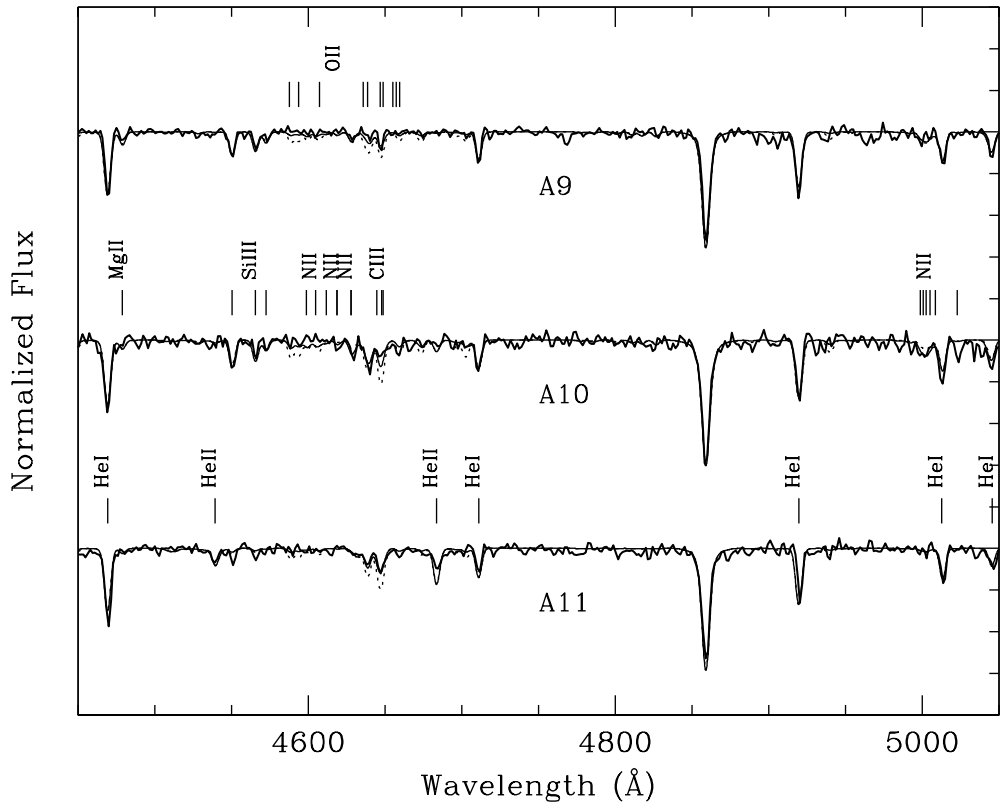}
\caption{As Fig.~\ref{models1}, for the wavelength range 4550--5050~\AA.}\label{models2}
\end{figure*}
\epsscale{1.0}

\epsscale{1}
\begin{figure*}
\plotone{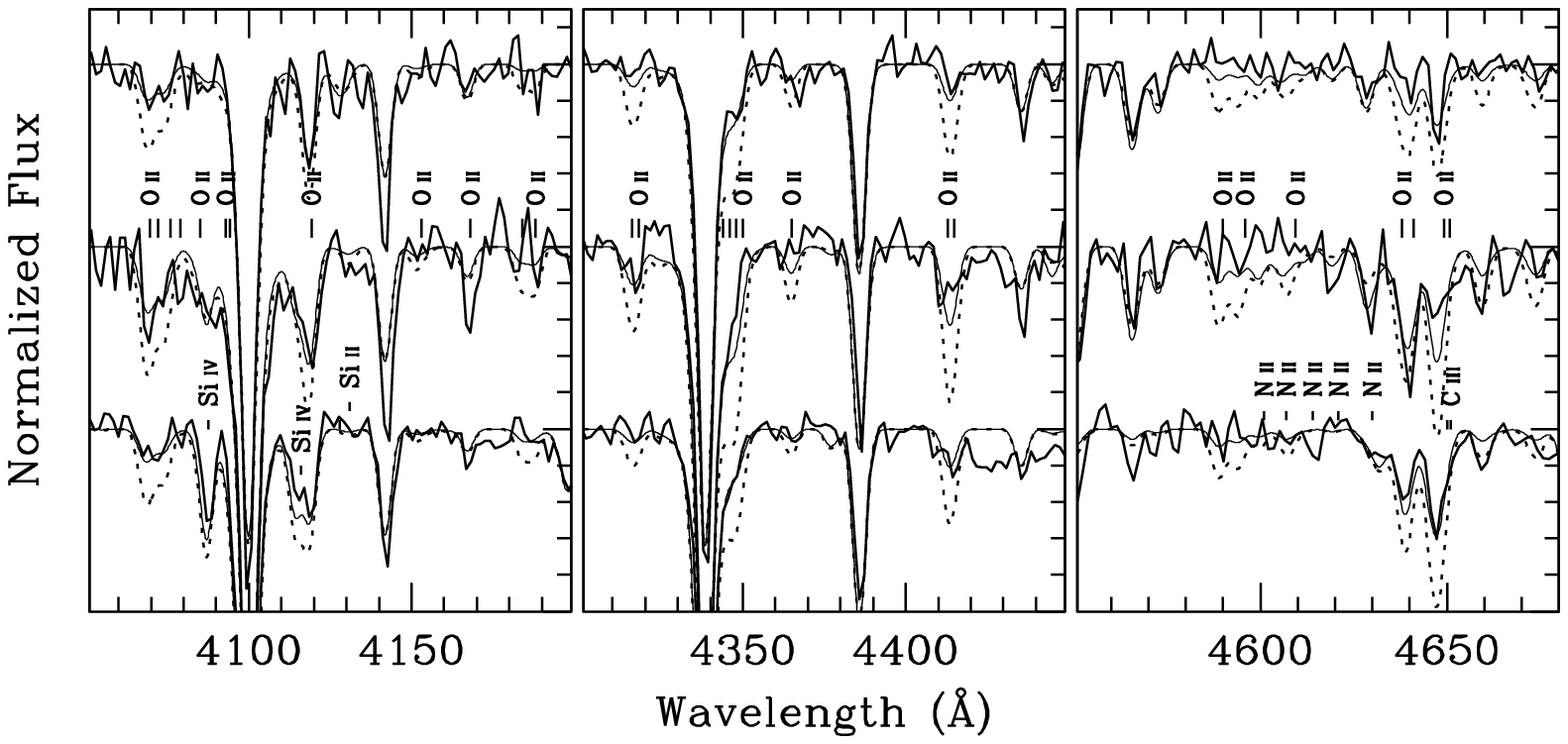}
\caption{Enlargment of three portions of the spectra and models presented in Fig.~\ref{models1}-\ref{models2}, zooming in on prominent \oii\/ features. The observed spectra of A9 (top), A10
(middle) and A11 (bottom) are shown with thick lines, the models calculated with our adopted 
oxygen abundances and 12\,+\,log(O/H)\,=\,8.45 are given by the thin and dashed lines, respectively.}\label{models_HR}
\end{figure*}
\epsscale{1.0}

\section{B-star abundances}\label{sec_abundances}
The case of WLM is peculiar among dwarf irregular galaxies of the Local Group, in that the stellar oxygen abundance obtained for one A-type supergiant (A14\,=\,SC~15) is about five times larger than measured in \hii\/ regions: [O/H]\,=\,$-0.21$ (\citealt{venn03}), corresponding to 12\,+\,log(O/H)\,=\,8.45 when adopting the solar value 12\,+\,log(O/H)$_\odot$\,=\,8.66 (\citealt{asplund04}).
The abundance of magnesium, also an $\alpha$-element, relative to the solar value determined from SC~15 and an additional A-type supergiant, SC~31 (=\,B12) is however 2.5 times lower than that of oxygen, [Mg/H]\,=\,$-0.62$. This is marginally consistent with the nebular oxygen abundance, [O/H]\,=\,$-0.83$ [12\,+\,log(O/H)\,=\,7.83] measured by \citet{lee05}.

Since both BA-type supergiants and \hii\/ regions are young  
objects, they should share a common chemical composition. The result found 
for oxygen in SC~15 is therefore puzzling.
While a few explanations for the discrepancy between the nebular and the stellar oxygen abundance observed in WLM have been proposed by Venn et al., it is desirable to
seek a comparison betweeen stellar and nebular abundances using additional 
stars, in particular early-B stars, whose spectra contain a large number of \oii\/ lines (for A supergiants the measurement of the oxygen abundance relies on 
a single line, \oi\,\lin6158).
We present here chemical abundances measured for three early-B supergiants, and we postpone the results for late-B and A stars to a 
future publication. 

We have focused on the three brightest O/early-B stars: A9 (=\,SC~35, B1.5~Ia), A10 (=\,SC~26, B0~Iab) and A11 (=\,SC30, O9.7~Ia).
The technique we employed for the abundance analysis of B supergiant spectra at FORS resolution has been explained in
detail elsewhere (\citealt{urbaneja03,urbaneja05n300}). Stellar parameters and metal abundances are derived via comparisons with model spectra calculated with the {\sc fastwind} code (\citealt{santolaya97}, \citealt{puls05}). The effective temperature is obtained from the silicon ionization balance (\siii/\siiii\/ and/or \siiii/\siiv)
or, in the case of the late-O star A11, the helium ionization balance (\hei/\heii).
The surface gravity is obtained from fits to the hydrogen Balmer lines.

In order to determine the abundance of metals the principal features we rely on are: \oii\llin4072-76, \oii\llin4317-19, \oii\llin4414-16, \nii\lin3995, \nii\lin5050 (blend), \cii\lin4267, \mgii\lin4481, \siii\llin4128-32, \siiii\llin\,4553-68-75 and \siiv\llin\,4089,4116. However, there are a large number of metal features present 
in the spectra of early-B stars, especially due to O and N. At low spectral resolution we cannot base our abundance estimates on single metal features, therefore the abundance solution is obtained by attempting to reproduce as many spectral features as possible with the model spectra.
This is especially critical for stars of low  metal content, as is the case here, because of the weakness of the lines. Our estimated uncertainties are 1000~K on \teff, 0.1 dex on log~$g$ and 0.2 dex on the metal abundances (for details see \citealt{urbaneja05n300}). Helium abundances are based on \hei\lin4026, \lin4388, \lin4471, \lin4921, \lin5015, \lin5048 and \lin6678. The additional helium lines present in the spectra are
not considered in the helium abundance calculation because of uncertainties in the corresponding
Stark broadening data. These lines are nevertheless included in the model fits presented below.

In Fig.~\ref{models1} and \ref{models2} we show with continuous thin lines our adopted fits to the spectra of A9, A10 and A11 (thick lines), in the wavelength intervals 3850-4450~\AA\/ and 4450-5050~\AA, respectively. 
In these figures the dotted lines represent {\sc fastwind} models in which 
the oxygen abundance was raised to the value measured by \citet{venn03}, 12\,+\,log(O/H)\,=\,8.45 in SC~15.
As these figures and the enlargements in Fig.~\ref{models_HR} show, such a high value is ruled out  by comparing the observed and calculated strengths of several oxygen lines and line blends, e.g.~those present in the red wing of H$\gamma$ at \lin4340~\AA, \oii\llin4072-76 and \oii\llin4414-16.

The derived stellar parameters and chemical abundances are summmarized
in Table~\ref{parameters}. For the parameters that are distance-dependent, we have assumed a distance to WLM of 0.93~Mpc ($m-M=24.85$, \citealt{mcconnachie05}). The reddening has been obtained from 
a comparison of the observed $V-I$ color index with the value derived from the spectral energy distribution
of the best-fit models. In all three cases $E(B-V)$ is consistent with a very small or a negligible 
component in addition to the foreground [$E(B-V)$\,=\,0.035, \citealt{schlegel98}].
The metal lines in these spectra are in general very weak, and only uppper limits on the abundances of  some elements can be provided, as in the case of
magnesium and carbon. For magnesium we obtained [Mg/H]\,$<$\,$-0.6$. The result from \citet{venn03} for two
A supergiants is [Mg/H]\,$=$\,$-0.62$.\\[-3mm]

\input{tab2}

{\em Oxygen.--} The oxygen abundance we measure for the three stars we have analyzed 
is in the range 12\,+\,log(O/H)\,=\,7.7--7.9. The spread is within our uncertainties.
Therefore, combining the three stars we derive a mean oxygen abundance 
12\,+\,log(O/H)\,=\,7.83 $\pm$ 0.12. This value agrees extremely well with the
findings of \citet{lee05}. Their average of direct, electron-temperature based
abundances for two \hii\/ regions yields 12\,+\,log(O/H)\,=\,7.83 $\pm$ 0.06.
Two of our stars, A9 and A10, are positioned within $\sim$1 arcmin (270 pc in projection) south of one of the two nebulae, HM9. The A11 supergiant is located further away, 360-450 pc
from both \hii\/ regions, HM9 and HM7, towards the east, but quite near ($\sim$\,70 pc in projection) the A-type supergiant analyzed in the SE region of WLM by \citet{venn03}, SC~31.
This proximity contributes to make the hypothesis of chemical inhomogeneity in this part of WLM less likely (see the discussion at the end of this section).\\[-3mm]

\begin{figure*}
\epsscale{0.85}
\plotone{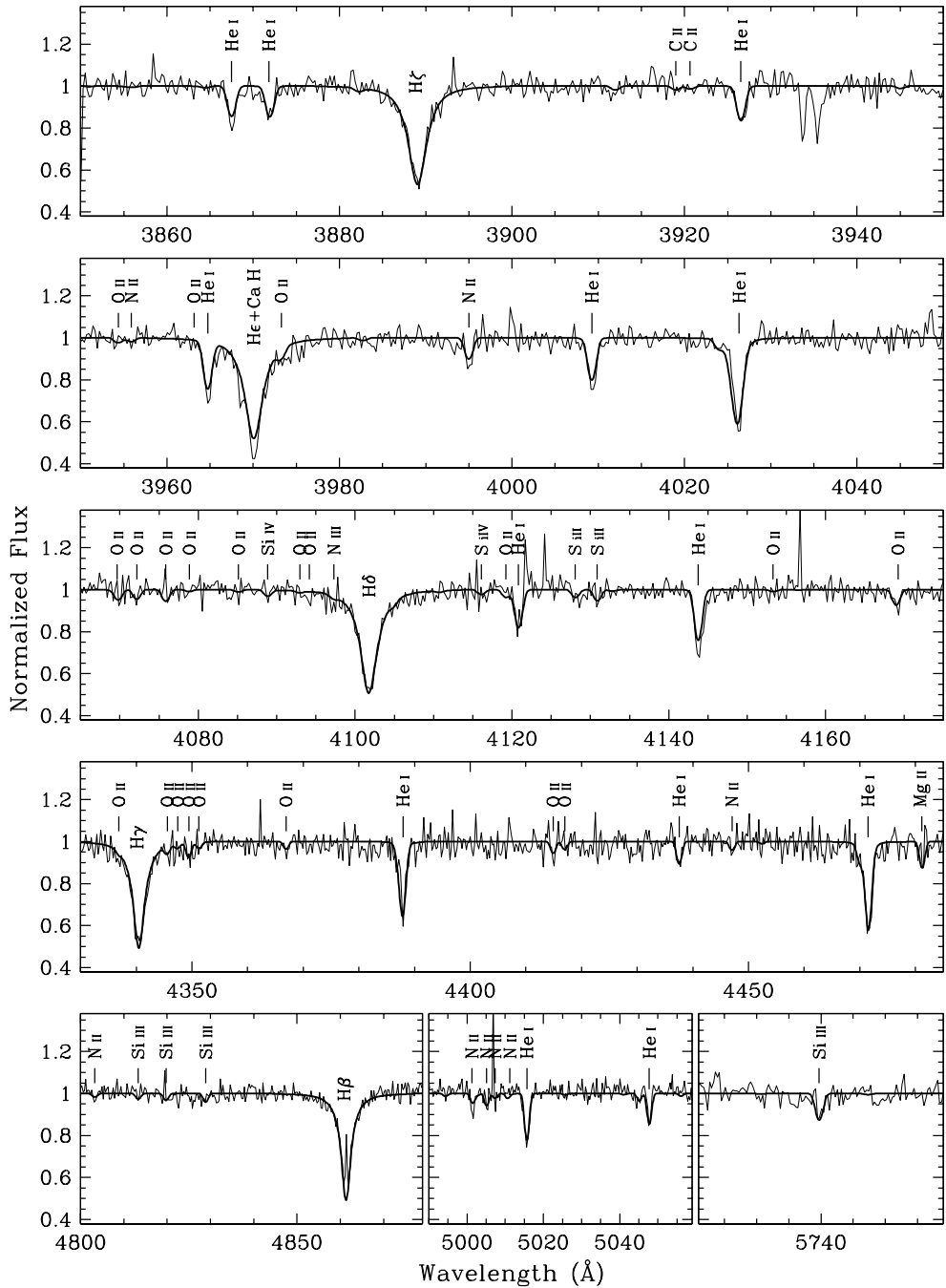}
\caption{The UVES spectrum of the B1.5\,Ia supergiant A9 (thin line) is
compared to a {\sc fastwind} model calculated with the stellar parameters 
and chemical abundances derived from the FORS spectrum, and summarized 
in Table~2. We show here the portions of the spectrum containing the 
most important metal features, labeled in the top half of each panel.
}\label{uves}
\end{figure*}
\epsscale{1.0}

{\em Nitrogen.--} The average abundance we measure is 12\,+\,log(N/H)\,$\simeq$\,7.5, the same 
value found by \citet{venn03} for SC\,15.
This is 1.2 dex higher than the average nitrogen abundance in the two \hii\/  regions
studied by \citet{lee05}. Considering the nebular value as the baseline for nitrogen in WLM, we therefore find an overabundance of nitrogen on the surface of the B-type supergiants of more 
than of a factor 10. 
From comparable overabundances observed in B-type supergiants in the SMC, \citet{trundle05} argue 
in support of the nitrogen enrichments   predicted at the end of the core hydrogen burning phase by stellar evolution models that include rotational mixing (\citealt{maeder01}).\\[-3mm]

{\em Silicon.--} Our measured silicon abundances, 12\,+\,log(N/H)\,=\,6.6--7.0 for the 3 different supergiants, are again in good agreement with the range of values determined in SC\,15 and SC\,31 by \citet{venn03}, 12\,+\,log(N/H)\,=\,6.57--6.95. Since silicon is an $\alpha$-process element, its abundance is expected to follow that of oxygen. In our case, [Si/O] varies within a relatively wide range, between $-0.14$ and 0.46. We note, however, that a large scatter in the silicon abundance is also observed in other B-type
supergiant samples, e.g.~in the SMC (\citealt{lennon03}, \citealt{trundle04}) and M33 (\citealt{urbaneja05m33}).

\subsection{UVES spectrum of A9}

A high-resolution ($R=32000$) spectrum was obtained at the ESO VLT with UVES by \citet{venn03} for one of the B supergiants we have analyzed (A9\,=\,SC35). This offered us the possibility of checking our abundance determinations for this star, and in particular to test whether the lower resolution of our FORS data affects the derivation of the chemical abundances. A 60-minute
exposure  resulted in a signal-to-noise ratio of about 50 per resolution
element in the UVES spectrum. The wavelength coverage extends from 3800\,\AA\/ to 10250\,\AA, 
but gaps are present around 4600\,\AA, 5600\,\AA\/ and in the red. As a result, 
the spectrum of A9 lacks the \siiii\llin4553-68-75 lines, which are crucial \teff\/ diagnostic in the early-B spectral domain. For this reason, \citet{venn03} were not able to measure chemical abundances for this B supergiant.

During our analysis we did not carry out an independent abundance study for this star. Instead, we 
verified that the model spectrum calculated for the FORS data with the parameters found in Table~\ref{parameters} provides a good match to the spectral lines in the UVES data. The comparison between our model and the data 
is presented in Fig.~\ref{uves}, where we show the portions of the spectrum
that contain the most important helium and metal features.
We did not find it necessary to adjust 
the chemical abundances, and in particular we confirmed the low $\alpha$-element 
content of A9. The result of this test is important, because it further strengthens our confidence in the abundances we are able to derive  from spectra obtained with a resolution of 4-5~\AA\/ (\citealt{przybilla06}).\\[-3mm]

In conclusion, our results obtained for three supergiants do not substantiate the oxygen abundance discrepancy found by  \cite{venn03} between \hii\/ regions and the A supergiant SC~15 (=\,A14). For other
chemical species in common (N, Si, Mg), instead, we find similar abundances.
The nebular and stellar oxygen abundances in WLM appear
to be in excellent agreement, as previously found in other dwarf irregular galaxies
of the Local Group, e.g.~NGC~6822 (\citealt{venn01}) and Sextans~A (\citealt{kaufer04}). This result confirms the peculiar nature of the
star analyzed by \citet{venn03}. Considering the different scenarios proposed by Venn et al.~to explain this peculiarity, \citet{lee05} noted that the origin for the
abundance discrepancy could be related to the location of SC~15 on the eastern side of the galaxy, where a peak in the \hi\/ distribution is found (\citealt{jackson04}), possibly connected
with the presence of \hii\/ regions with enhanced extinction. This suggested  
unusual processes in this part of WLM (for example, dilution of the nebular 
chemical composition by metal-poor gas infall).
However, we note that the SC~30 (=\,A11) supergiant analyzed in the current paper also lies
in the same eastern region of the galaxy. From the B-star data presented here and from the available nebular abundances we do not find support for the presence
of chemical spatial inhomogeneities in WLM.

\section{Photometric variability}

\begin{figure}
\plotone{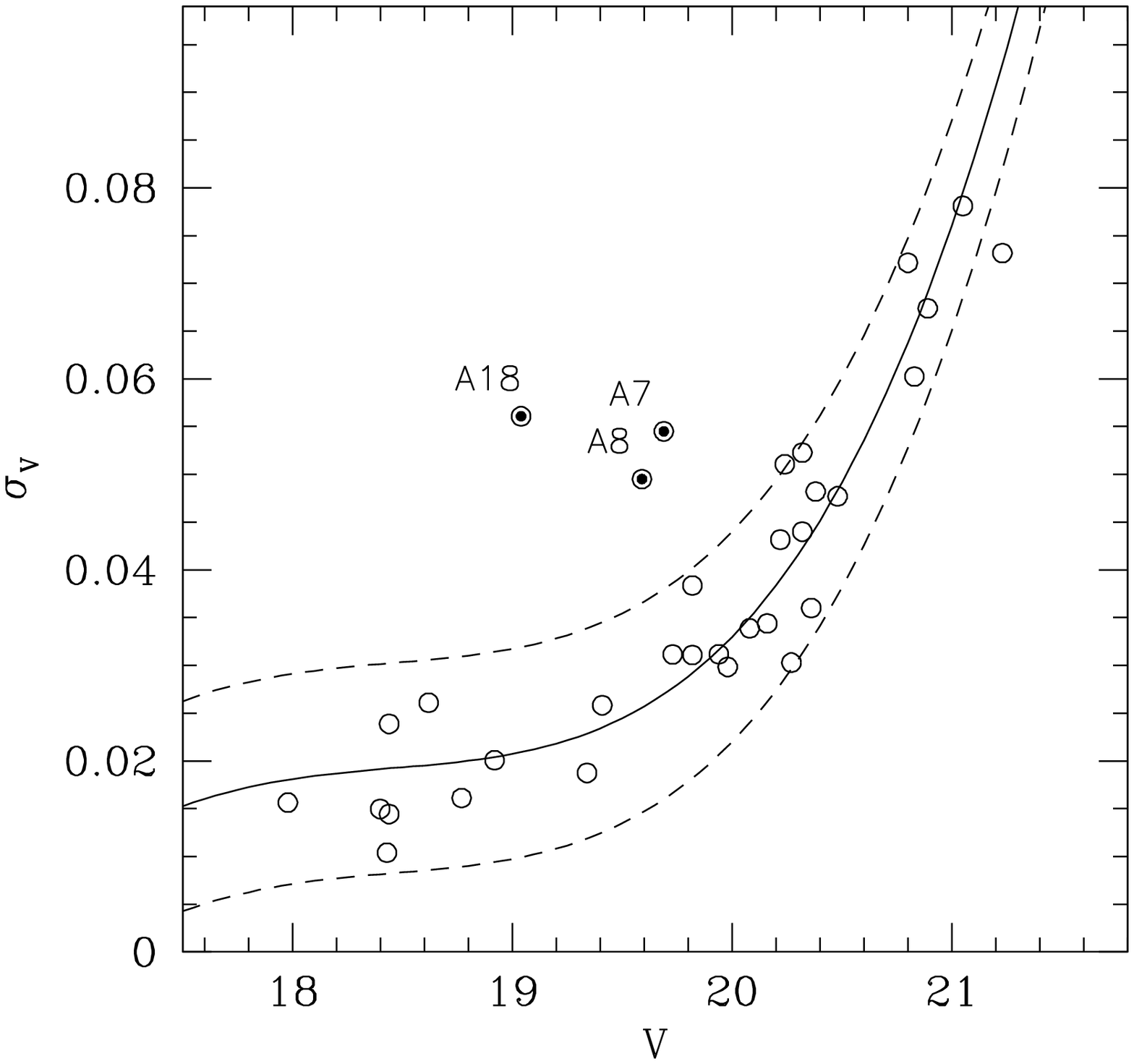}
\caption{The standard deviation in the $V$ band, $\sigma_V$, from measurements at  
approximately 100 epochs, is shown as a function of $V$ magnitude for 33 of the spectroscopic targets (open circles). The trend of the median value for nearly 5000 stars in WLM is shown with the continuous line. The dashed lines are drawn 
at the approximate 1-$\sigma$ deviation from the median (0.11 mag). The three
outliers A7, A8 and A18 are indicated with the different symbols.
}\label{variability}
\end{figure}

In our work on NGC~300 we found that about 20\% of the blue supergiants we investigated spectroscopically in that galaxy are variable (\citealt{bresolin04b}). However, we concluded from the analysis of 14 B8-A2 supergiants with light amplitudes in the $V$ band between 0.08 mag and 0.23 mag that the observed variability does not significantly affect the \fglr\/ (\citealt{kudritzki03}).
We have examined our multi-epoch photometry of the spectroscopic targets in WLM in order to verify the presence of variability. For most stars we have $V$ photometry for approximately 100 epochs, extending over a period of more than two years (October~2001--December~2003). We have compared the standard deviation $\sigma_V$ of the
$V$ magnitudes of the spectroscopic targets with that of the general stellar population in WLM, as shown in Fig.~\ref{variability}. Most of the stars for which we obtained spectra have $\sigma_V$ that are consistent with the observational scatter of non-variable stars. Only three outliers 
are found: A18 (spectral type G2), A7 (B1.5~Ia) and A8 (B3~Ib), with $\sigma_V\simeq0.05$ mag. The variability of these three stars is significant at the 2-3$\sigma$ value. The analysis of their
light curves did not reveal any significant periodicity, however. The star A18 (=\,SC 4) was identified as 
a red irregular variable by \citet{sandage85}. In fact, it is the one with the largest amplitude in the $B$-band (0.69 mag) in their sample.

We conclude that most of the spectroscopic targets are non-variable stars, at least within the limits of our photometric acccuracy. The absence of variability at the
0.1 mag level or more for all the spectroscopic targets implies that even in the case of WLM the photometric variability is not a cause for concern in the determination of the \fglr.

\section{Summary and conclusions}

As part of our photometric and spectroscopic investigation of the stellar content of nearby galaxies, the Araucaria Project, we have presented the first extensive spectral catalog of stars in the dwarf irregular galaxy WLM.  In our series of papers focused on the analysis of blue supergiants, our first step involves the classification of the available spectra, since it provides a first order approach to the physical characteristics of the target stars.
In the case of WLM, as well as in additional dwarf irregulars that are part of the project and that will be considered in future publications (e.g.~IC~1613 and NGC~6822), this process needs to account for the low metal content of the galaxy relative to the galactic standards used in the MK process. We have therefore based our
spectral classification on the criteria developed for B and  supergiants contained in the Small Magellanic Cloud
by \citet{lennon97} and \citet{evans03}, respectively. 

Our spectral catalog, presented in Table~1 and in Fig.~\ref{spectraA1}-\ref{spectraB}, shows that 
our higher S/N spectroscopic sample of 19 objects, selected from $VI$ photometry, contains at least 6 early-B (B0-B5) supergiants and 6 late-B and early-A (B8-A2) stars of luminosity class between Ia and II, as well as an O7~V star and an O9.7~Ia star. The spectra of several of these stars is of sufficient quality for 
a determination of the stellar parameters and abundances.
We have acquired also a second set of lower S/N spectra for mostly BA stars, however their quality in general 
does not allow us to carry out further analysis.

We have carried out a quantitative analysis for three of our targets: A9 (B1.5~Ia), A10 
(B0~Iab) and A11 (O9.7~Ia). Using {\sc fastwind} models we have derived chemical abundances 
for oxygen, nitrogen and silicon, while upper limits were estimated for carbon and magnesium.
The oxygen abundance for the three supergiants is consistent with a single value corresponding to
12\,+\,log(O/H)\,=\,7.83 $\pm$ 0.12. This oxygen abundance is in excellent agreement with the 
measurement derived from \hii\/ regions. Our one B-supergiant spectrum in the SE region does not confirm the high oxygen abundance from the A-type supergiant in this region. From our B-supergiant analysis
the SE region could thus have the same oxygen abundance as the rest of WLM.

In addition to the measurement of chemical abundances in blue supergiants, we are focusing on the use of these stars as extragalactic distance indicators, through the Flux-weighted Gravity--Luminosity Relationship ({\sc fglr}). The \fglr\/ for supergiants in WLM will require the determination of stellar parameters for a larger sample of objects, and will be presented in an upcoming publication. Here we have shown that a possible complication in the \fglr\/ method, namely the photometric variability of the target stars, is negligible, confirming our previous conclusion (\citealt{bresolin04b}) that the \fglr\/ is insensitive to the microvariability of blue supergiants.\\

\acknowledgments We thank L. Rizzi for help
in the astrometric calibration of the FORS2 WLM image.
GP and WG gratefully aknowledge financial support from the Chilean Center for Astrophysics
FONDAP 15010001. 

\bibliography{wlm}

\clearpage

\clearpage



\clearpage


\end{document}

%% file: tab1.tex
\begin{deluxetable*}{rrrrrlccccl}
\tabletypesize{\scriptsize}
\tablecolumns{11}
\tablewidth{0pt}
\tablecaption{WLM - Spectroscopic targets}

\tablehead{
\colhead{Slit}     &
\colhead{R.A.}            &
\colhead{Decl.}           &
\colhead{}             &
\colhead{}     &
\colhead{Spectral}    &
\colhead{}			& 
\colhead{} &
\colhead{} &
\colhead{$V_{\rm Helio}$} &
\colhead{}\\
\colhead{number}   &
\colhead{(J2000)}         &
\colhead{(J2000)}         &
\colhead{$V$}         &
\colhead{$V-I$}         &
\colhead{Type}		&
\colhead{K/(H + H$\epsilon$) }			&
\colhead{$W_\gamma$}			&
\colhead{S/N}	&
\colhead{(km/s)}	&
\colhead{Comments}\\[1mm]
\colhead{(1)}	&
\colhead{(2)}	&
\colhead{(3)}	&
\colhead{(4)}	&
\colhead{(5)}	&
\colhead{(6)}	&
\colhead{(7)}	&
\colhead{(8)}	&
\colhead{(9)}	&
\colhead{(10)}	&
\colhead{(11)}}
\startdata
\\[-1mm]
Set A & & & & & & & & & &\\
1\dotfill   & 0:01:58.46 &	$-$15:24:33.5 &     20.89 & 1.50 & G5~I       & $\sim1$     & $\sim3$    &  16 & $-187$ & \\
2\dotfill   & 0:01:59.04 &	$-$15:24:42.6 &     20.16 & 0.09 & A0~II      & 0.27        & 6.0        &  44 & $-186$ & SC~68\\
3\dotfill   & 0:01:56.63 &	$-$15:25:01.2 &     21.05 & 1.15 & G0~I       & 0.85        & 2.2        &  18 & $-168$ & \\
4\dotfill   & 0:02:01.59 &	$-$15:25:26.9 &     20.22 & 0.07 & A2~II      & 0.41        & 7.1        &  44 & $-163$ & \\
5\dotfill   & 0:02:03.32 &	$-$15:25:52.5 &     19.41 & $-$0.04 & B8~Iab  & 0.17        & 2.9        &  64 & $-148$ & SC~45 \\
6\dotfill   & 0:01:56.16 &	$-$15:26:24.6 &     19.82 &  0.38 & A7~Ib     & 0.94        & 7.2        &  49 & $-136$ & SC~51\\
7\dotfill   & 0:01:58.12 &	$-$15:26:48.6 &     19.69 & $-$0.18 & B1.5~Ia & 0.35        & 1.6        &  58 & $-135$ & SC~37 \\
8\dotfill   & 0:01:55.04 & 	$-$15:26:59.9 &     19.59 & $-$0.05 & B3~Ib   & 0.31        & 3.0        &  60 & $-143$ & SC~55 \\
9\dotfill   & 0:01:57.21 &	$-$15:27:18.1 &     18.44 & $-$0.06 & B1.5~Ia & 0.28        & 1.6        & 101 & $-128$ & SC~35 \\
10\dotfill  & 0:01:54.08 &	$-$15:27:45.7 &     19.34 & $-$0.15 & B0~Iab  & 0.19        & 2.3        &  68 & $-126$  & SC~26\\
11\dotfill  & 0:01:59.97 &	$-$15:28:19.2 &     18.40 & $-$0.18 & O9.7~Ia & 0.25        & 1.7        & 106 & $-123$  & SC~30\\
12\dotfill  & 0:01:53.24 &	$-$15:28:39.7 &     17.98 & 0.06 & B9~Ia      & 0.26        & 1.9        & 119 & $-118$ & SC~22\\
13\dotfill  & 0:01:53.35 &	$-$15:28:52.2 &     19.94 & $-$0.15 & B2.5~Ib & 0.18        & 2.7        &  52 & $-113$  & SC~21\\
14\dotfill  & 0:01:59.58 &	$-$15:29:26.3 &     18.43 & 0.23 & A2~II      & 0.52        & 6.2        &  96 & $-94$  & SC~15\\
15\dotfill  & 0:02:00.53 &	$-$15:29:52.1 &     20.36 & $-$0.22 & O7~V((f)) & 0.16      & 2.2        &  47 & $-115$ & SC~14\\
16\dotfill  & 0:01:57.90 &	$-$15:30:13.4 &     18.44 & 0.16 & A2~Ia      & 0.35        & 2.4        &  96 & $-68$  & SC~10\\
17\dotfill  & 0:02:00.83 &	$-$15:30:25.0 &     19.34 & 0.00 & B5~Ib      & 0.19        & 2.8        &  67 & $-74$  & SC~9\\
18\dotfill  & 0:01:59.63 &	$-$15:31:00.3 &     19.04 & 1.69 & G2~I       & $\sim1$     & 2.0        &  30 & $-73$  & SC~4\\
19\dotfill  & 0:02:00.84 &	$-$15:31:16.0 &     18.62 & 1.62 & G2~I       & 1.4         & 2.0        &  36 & $-90$  & SC~6\\[4mm]
Set B & & & & & & & & & &  \\
1\dotfill   & 0:02:05.16 &	$-$15:24:22.7	 &   21.00 & 	0.37 & A2~II & 0.38        & 7.6        & 21  & $-162$ & \\
2\dotfill   & 0:02:04.38 &	$-$15:24:46.3	 &   20.38 &  	0.06 & A2~Ia & 0.38        & 2.0        & 27  & $-145$ & \\
3\dotfill   & 0:01:55.69 &	$-$15:24:48.8	 &   20.48 &  	0.01 & B5~II & 0.24        & 3.6        & 28  & $-125$ & SC~58\\
4\dotfill   & 0:01:59.89 &	$-$15:25:28.0	 &   20.83 & 	0.36 & A5~II & 0.81        & 10.0       & 19  & $-154$ & \\
5\dotfill   & 0:02:00.03 &	$-$15:25:44.8	 &   20.80 &  	0.08 & A0~II & 0.25        & 5.0        & 23  & $-120$ & \\
6\dotfill   & 0:01:56.37 &	$-$15:26:06.5	 &   20.02 & 	0.28 & A2~II & 0.37        & 8.7        & 29  & $-131$ & SC~52\\
7\dotfill   & 0:01:57.17 &	$-$15:26:14.3	 &   20.32 & 	0.32 & A3~II & 0.56        & 7.0        & 25  & $-128$ & \\
8\dotfill   & 0:01:56.75 &	$-$15:26:36.6	 &   20.24 & 	0.22 & A3~II & 0.65        & 6.5        & 28  & $-136$  & \\
9\dotfill   & 0:02:01.93 &	$-$15:27:25.3	 &   19.82 &  	0.09 & A0~II & 0.27        & 5.5        & 35  & $-104$ & SC~42\\
10\dotfill  & 0:01:53.21 &	$-$15:27:29.2	 &   19.04 &	 $-$0.12 & B2~II & 0.29        & 3.2        & 53  & $-103$  & SC~27\\
11\dotfill  & 0:01:56.35 &	$-$15:27:58.4	 &   19.75 & 	0.17 & A0~Iab & 0.31        & 3.4        & 32  & $-108$ & \\
12\dotfill  & 0:02:00.64 &	$-$15:28:29.9	 &   18.77 & 	0.26 & A2~II & 0.51        & 6.6        & 57  & $-118$ & SC~31\\
13\dotfill  & 0:01:53.65 &	$-$15:28:29.9	 &   18.92 &	 $-$0.14 & B1~Ia & 0.20        & 1.8        & 57  & $-101$  & SC~23\\
14\dotfill  & 0:01:56.46 &	$-$15:29:01.7	 &   19.98 &  	0.03 & A0~Ib & 0.19        & 4.2        & 32  & $-89$  & \\
15\dotfill  & 0:02:00.05 &	$-$15:29:31.1	 &   20.32 & 	0.36 & A5~II & 0.77        & 7.2        & 28  & $-98$  & \\
16\dotfill  & 0:01:58.75 &	$-$15:30:01.7	 &   20.08 & 	0.20 & composite & 0.15    & 4.7        & 29  & $-93$  & SC~12\\
17\dotfill  & 0:02:00.21 &	$-$15:30:14.2	 &   20.27 & 	0.16 & composite & $<$0.08 & 2.2        & 26  & $-81$ & \\
18\dotfill  & 0:01:59.44 &	$-$15:30:46.9	 &   21.23 & 	0.18 & composite & \nodata & \nodata    & 17  & \nodata & \\
19\dotfill  & 0:02:00.50 &	$-$15:31:08.4	 &   19.73 & $-$0.12 & B2~Ib  & 0.21        & 2.9       & 37   & $-87$  & \\
\enddata
\end{deluxetable*}

%% file: tab2.tex
\begin{deluxetable}{cccc}
\tabletypesize{\scriptsize}
\tablecolumns{4}
\tablewidth{0pt}
\tablecaption{B supergiants - Best fit parameters\label{parameters}}

\tablehead{
\colhead{\phantom{}Properties\phantom{abcdefg}}	 &
\colhead{A9}     		&
\colhead{A10}            &
\colhead{A11}           }
\startdata
\\[-1mm]
Spectral type \dotfill							&  B1.5~Ia			&  B0~Iab    			&  O9.7~Ia  \\
\teff\/ (K)\dotfill								&  20000 $\pm$ 1000 	&  25000 $\pm$ 1000		&  29000 $\pm$ 1000   \\
log $g$ (cgs)\dotfill							&  2.45 $\pm$ 0.10  	&  2.90 $\pm$ 0.10     	&  3.00 $\pm$ 0.10    \\
$R/R_\odot$\dotfill								&  39 $\pm$ 2    	&  20 $\pm$ 2      		&  26 $\pm$ 2     \\
B.C.\dotfill				    						&  $-1.91$			&  $-2.46$             	&  $-2.83$            \\
M$_{\rm bol}$\dotfill		       				&  $-8.58\pm0.13$   	&  $-8.15\pm0.14$      	&  $-9.33\pm0.13$     \\
log $L/L_\odot$\dotfill		                    	&  $5.33\pm0.14$    	&  $5.16\pm0.14$       	&  $5.63\pm0.13$      \\
$M^{\rm spec}/M_\odot$\dotfill                   	&  $14\pm6$         	&  $11\pm5$            	&  $23\pm9$          \\
$E(B-V)$	\dotfill									 &  0.06              &  0.04                 &  0.04      \\
$\upsilon_{\rm turb}$ (km\,s$^{-1}$)\dotfill     	&  12  				&  15  					&  15 \\
$Y_{\rm He}$\dotfill								&  0.12             	&  0.12                	&  0.15  \\
$\epsilon_{\rm C}$\dotfill  						&   7.4          	&  $<6.9$                	&  $<6.9$ \\
$\epsilon_{\rm N}$\dotfill  						&   7.4            	&  7.8                	&  7.4 \\
$\epsilon_{\rm O}$\dotfill  						&   7.7            	&  7.9                	&  7.9 \\
$\epsilon_{\rm Mg}$\dotfill 						&   $<7.0$         	&  $<7.0$             	&  $<7.0$ \\
$\epsilon_{\rm Si}$\dotfill 						&   7.0            	&  7.0                	&  6.6 \\
$[{\rm O/H}]$ (dex)\dotfill						&  $-1.0$    		&  $-0.8$              	&  $-0.8$\\
$[{\rm Mg/H}]$ (dex)\dotfill				     &  $<-0.6$           &  $<-0.6$               &  $<-0.6$\\
$[{\rm Si/O}]$ (dex)\dotfill						&  $0.46$    		&  $0.26$              	&  $-0.14$\\

\enddata

\end{deluxetable}